\documentclass[a4paper,fleqn]{cas-dc}
\usepackage[round]{natbib}

\usepackage{eurosym,makecell,comment}
\usepackage{multirow}
\usepackage{threeparttable}
\usepackage{framed}
\usepackage{hyperref}
\usepackage{url}
\usepackage{verbatim}
\usepackage{graphicx}
\usepackage{subcaption}
\usepackage{float}
\usepackage{xcolor}
\usepackage{microtype}
\usepackage{colortbl}
\usepackage{amsmath,amssymb,amsfonts}
\usepackage{algorithmic}
\usepackage{algorithm}
\usepackage{mathtools}

\usepackage{array} 

\usepackage{etoolbox}
\usepackage{soul} 

\newtoggle{finalPaper}

\setstcolor{red}

\toggletrue{finalPaper} 

\iftoggle{finalPaper} {

	\newcommand{\rmvtxt}[1]{}}
{

	\newcommand{\rmvtxt}[1]{\st{#1}}}

\begin{document}
\let\WriteBookmarks\relax
\def\floatpagepagefraction{1}
\def\textpagefraction{.001}
\shorttitle{Neuronal Jamming Cyberattack over Invasive BCI Affecting the Resolution of Tasks Requiring Visual Capabilities}
\shortauthors{L\'opez Bernal et~al.}

\title[mode = title]{Neuronal Jamming Cyberattack Over Invasive BCI Affecting the Resolution of Tasks Requiring Visual Capabilities}

\author[1]{Sergio L\'opez Bernal}[orcid=0000-0003-1869-1965]
\cormark[1]

\author[2]{Alberto Huertas Celdr\'an}[orcid=0000-0001-7125-1710]

\author[1]{Gregorio Mart\'inez P\'erez}[orcid=0000-0001-5532-6604]

\address[1]{Department of Information and Communications Engineering, University of Murcia, Murcia 30100, Spain}

\address[2]{Communication Systems Group (CSG), Department of Informatics (IfI), University of Zurich UZH, 8050 Zürich, Switzerland}

\cortext[cor1]{Corresponding author.
Email address: slopez@um.es (S. L. Bernal)}

\begin{keywords}
Cybersecurity \sep Safety \sep Neuronal Cyberattacks \sep Convolutional Neural Networks \sep Brain-Computer Interfaces

\end{keywords}

\maketitle      

\begin{abstract}
Invasive Brain-Computer Interfaces (BCI) are extensively used in medical application scenarios to record, stimulate, or inhibit neural activity with different purposes. An example is the stimulation of some brain areas to reduce the effects generated by Parkinson's disease. Despite the advances in recent years, cybersecurity on BCI is an open challenge since attackers can exploit the vulnerabilities of invasive BCIs to induce malicious stimulation or treatment disruption, affecting neuronal activity. In this work, we design and implement a novel neuronal cyberattack, called Neuronal Jamming (JAM), which prevents neurons from producing spikes. To implement and measure the JAM impact, and due to the lack of realistic neuronal topologies in mammalians, we have defined a use case with a Convolutional Neural Network (CNN) trained to allow a mouse to exit a particular maze. The resulting model has been translated to a neural topology, simulating a portion of a mouse's visual cortex. The impact of JAM on both biological and artificial networks is measured, analyzing how the attacks can both disrupt the spontaneous neural signaling and the mouse's capacity to exit the maze. Besides, we compare the impacts of both JAM and FLO (an existing neural cyberattack) demonstrating that JAM generates a higher impact in terms of neuronal spike rate. Finally, we discuss on whether and how JAM and FLO attacks could induce the effects of neurodegenerative diseases if the implanted BCI had a comprehensive electrode coverage of the targeted brain regions.

\end{abstract}

\section{Introduction}
\label{sec:intro}

Brain-Computer Interfaces (BCIs) are devices providing bidirectional communication channels between the brain and external devices. One of the primary uses of BCI technologies is in health scenarios, where clinicians acquire relevant information about the brain for diagnosis purposes \cite{Lebedev:BrainMachineIF:2017}. Additionally, BCI systems enable artificial stimulation and inhibition of neuronal activity \cite{yao:stimulation:2019}. In particular, neurostimulation has been used in a wide variety of medical scenarios, ranging from the treatment of neurodegenerative diseases, such as Alzheimer's or depression \cite{kaufman:stimuli:2013}, to provide prosthetic users with feedback \cite{doherty:nature:2011}. Within these systems, there are two main categories based on their invasiveness. Non-invasive BCIs can externally stimulate the brain without surgery and, although some technologies can target small areas of the brain, non-invasive BCIs cover larger regions of the brain. In contrast, invasive systems can be applied to small areas, even with a single-neuron resolution, but introducing higher physiological risks \cite{ramadan:controlSignalsReview:2017}. 

Based on the relevance and expansion of BCIs, new technologies and companies have emerged in recent years, focusing on developing new invasive systems to stimulate the brain with neuronal granularity. This is the case of Neuralink \cite{Musk2019}, a company that has designed disruptive BCI systems to record data at the neuronal level, and it is currently working on covering the stimulation functionality. Besides, Neural Dust \cite{seo:neuralDust:2013} is an architecture of millions of nanoscale implantable devices, located in the cortex, that allow neural recording. An evolution of Neural Dust is the Wireless Optogenetic Nanonetworking device (WiOptND) \cite{Wirdatmadja:optogeneticModelProtocols:2017}, which uses optogenetics to stimulate the neurons. Although these approaches are promising, the authors of \cite{Lopez_Bernal:cyberattacks_implants:2020} have shown that they have vulnerabilities that could allow attackers to control both systems and perform malicious stimulation actions, altering spontaneous neuronal signaling. In the same direction, in \cite{Lopez_Bernal:cyberBCI:2021} was identified that the field of cybersecurity in BCI is not mature enough, and non-sophisticated attacks can generate significant damage. In summary, the BCI vulnerabilities could be exploited by attackers to take advantage of these promising neurostimulation technologies. Based on these works, this manuscript focuses on the scarce research dealing with cyberattacks aiming to alter neuronal behavior. Additionally, new ways to measure and understand the impact of these attacks are also required. In particular, these issues gain special relevance due to the possibility of attacks being able to worsen or recreate the effects of common neurodegenerative diseases \cite{Lopez_Bernal:cyberBCI:2021}.

Intending to improve the previous challenges, the main contribution of this work is the definition and implementation of a novel neuronal cyberattack, \textit{Neuronal Jamming cyberattacks} (JAM), focused on the inhibition of neural activity. The present work aims to explore the impact that inhibitory neuronal cyberattacks can generate over the brain. Nevertheless, there is an absence in the literature of comprehensive neuronal topologies and therefore, we simulate a portion of the visual cortex of mice, defining a use case of a mouse trying to exit a given maze. The neuronal topology has been built by using a Convolutional Neural Network (CNN) \cite{Geron2019} trained to solve this particular use case. The second contribution of this work is the evaluation of the impact caused by JAM cyberattacks over both neuronal and artificial simulation in this specific scenario. To perform the analysis we have used existing metrics but also defined a subset of new ones, concluding that JAM cyberattacks can alter spontaneous neuronal behavior and force the mouse to perform erratic decisions to escape the maze. The third main contribution of this work is to compare the impact caused by JAM with an existing cyberattack named Neuronal Flooding (FLO), from the biological and artificial perspectives. We have observed that applying a FLO cyberattack over the last positions of the maze generates a reduction of its effectiveness from both biological and artificial approaches. Additionally, JAM cyberattacks are more damaging when increasing the number of consecutive positions under attack, translated into a reduction in the neural activity and an augmentation in the number of steps to find the exit. From the comparison between scenarios, FLO presents a high Pearson correlation between experiments, of around 0.8, indicating a strong relationship. On its side, JAM presents worse results, which can be explained due to the particular restrictions during the implementation. Finally, we discuss the relationship that recent neuronal cyberattacks could have with neurodegenerative diseases. 

The remainder of the paper is structured as follows. Section~\ref{sec:related} reviews the state of the art in cybersecurity oriented to BCI and neuronal cyberattacks. After that, Section~\ref{sec:jamming} introduces the definition of the Neuronal Jamming cyberattack. Section~\ref{sec:setup} presents the experimental setup required to implement both JAM and FLO neuronal cyberattacks. Additionally, Section~\ref{sec:resultsJAM} and Section~\ref{sec:resultsFLO} describe, respectively, the results obtained after implementing JAM and FLO cyberattacks over multiple positions of the maze and the impact that they cause. These two sections also include a comparison of the relationship between artificial and biological approaches. Subsequently, Section~\ref{sec:discussion} discusses about the impact that neuronal cyberattacks can have on neurodegenerative diseases. Finally, Section~\ref{sec:conclusion} presents conclusions and future work. 

\section{Related Work}
\label{sec:related}

Cybersecurity applied to BCI is relatively recent, emerging in the last five years concepts such as brain-hacking or neurosecurity \cite{ienca:neuroprivacy:2015, ienca:hackingBrain:2016}. These publications identify that neurostimulation BCI devices present a high risk in patients' safety since an attacker could disrupt the treatment parameters. Additionally, they highlighted that attacks do not need to be complex to cause brain damage. 

During these recent years, the academic literature has widely focused on the study of cybersecurity in health scenarios, aiming to preserve patients' privacy or improving the security of clinical devices \cite{Huertas2017, Huertas2018}. However, the literature has focused on particular cybersecurity aspects of BCI, mostly from theoretical and ethical perspectives. Although previous studies have highlighted the applicability of cryptographic and jamming attacks \cite{ienca:hackingBrain:2016}, malware strategies \cite{bonaci:appStores:2015}, or potential attacks over BCI architectures \cite{ballarin:cybersecurityAnalysis:2018}, these works are scarce and focus on particular privacy and security aspects, not addressing the physical safety dimension. Additionally, the authors of \cite{takabi:privacyThreatsCounter:2016, bonaci:appStores:2015} identified that the platforms and frameworks used for the development of BCI applications could be vulnerable against cyberattacks. Based on that, the authors of \cite{Lopez_Bernal:cyberBCI:2021} performed a review of the state of the art in cybersecurity on BCI with a comprehensive analysis of physical safety issues, compiling already documented attacks over the BCI life-cycle, their impacts, and the countermeasures to detect and mitigate them. This work also studied the literature concerning attacks, impacts, and countermeasures from existing and prospecting architectural BCI deployments. Furthermore, they proposed the application of well-known attacks, impacts, and countermeasures from the cybersecurity domain to BCI. In a nutshell, they identified an enormous absence of works addressing cybersecurity aspects in BCI technologies. 

Regarding cyberattacks altering the behavior of neurons, the authors of \cite{Lopez_Bernal:cyberattacks_implants:2020} detected vulnerabilities in emerging neurostimulation technologies and defined two \textit{neuronal cyberattacks}, Neuronal Flooding (FLO) and Neuronal Scanning (SCA), aiming to disrupt the spontaneous behavior of the targeted zones of the brain. The FLO cyberattack consists in attacking, in a particular instant, a subset of neurons from the brain, while SCA targets one neuron per time instant, imitating the port scanning technique. They also defined several metrics to measure the impact of these attacks compared to the spontaneous neuronal activity. In short, they identified that both neuronal cyberattacks induced a considerable alteration in the spontaneous neural signaling.

The neuronal cyberattacks presented in \cite{Lopez_Bernal:cyberattacks_implants:2020} demonstrate the feasibility of performing attacks over the brain aiming to disrupt its spontaneous neural activity. However, they do not explore the physiological or psychological consequences that an alteration in neural signaling can generate. In that direction, the authors of \cite{Lopez_Bernal:cyberBCI:2021} theoretically proposed the feasibility of recreating the effect of neurodegenerative disorders such as Parkinson's and Alzheimer's diseases. For that, the neurostimulation system used would be required to cover the brain regions that are naturally impacted by these diseases and present vulnerabilities that can be exploited by attackers. This work highlighted the high impact that recreating neurodegenerative disorders could have on users' physical safety.

To understand how cyberattacks could affect the brain and its relationship with degenerative diseases, it is essential to mention that, from a neurological point of view, most brain disorders are revealed as a dysfunction of communication between neurons or with other organs, defining the term of \textit{brain connectivity disorders}. Within this term, we can include neurodegenerative diseases. Alzheimer's Disease (AD) is a progressive neurodegenerative disorder that induces the degradation and death of brain cells. It seems that neurodegenerative diseases spread along structurally connected neural networks, known as \textit{neuronal circuits}, presenting a functional relevance. There is a relationship between AD and changes in neuronal activity in the Default Mode Network circuit (DMN), where parts of the DMN present increased connectivity at the beginning of the disease, indicating compensation for the failure of other regions of the circuit before they degenerate. During the progression of AD, the deactivation of the DMN is gradually more pronounced. Nevertheless, it is not clear if the circuit disruption is a cause or a consequence of the disease \cite{Zott:Alzheimer:2018}.

Amyotrophic lateral sclerosis (ALS) is a neurodegenerative disease affecting cortical and spinal neurons, which generates a loss of muscle control and paralysis. ALS is associated with a dysfunction of cortical circuits, based on hyperexcitability of neuronal activity. Hyperexcitability can be understood as an exaggerated response to a stimulus, or the response to stimuli that generally do not induce a response. In this sense, ALS presents a perturbation in the excitatory/inhibitory balance, leading to pathological changes in cortical excitability \cite{Brunet:ALS:2020}. 

Despite the current knowledge about the behavior of neurodegenerative diseases, such as AD or ALS, there are no proposed cyberattacks in the literature trying to emulate the neuronal behavior of these conditions. Because of that, the current manuscript explores the possibility of inducing excitatory and inhibitory neuronal behavior to recreate these conditions in the long term.

\section{Neuronal Jamming Cyberattack}
\label{sec:jamming}

This section presents the formal definition of the Neuronal Jamming cyberattack (JAM), including algorithmic and graphical representations to ease its understanding.

Jamming is a well-known cyberattack aiming to block the legitimate communication between elements of a system using malicious interference. The result of this attack generates a denial of service (DoS) over the communication. From a neurological perspective, we conceive a jamming cyberattack as an inhibition of the spontaneous activity of a set of neurons during a particular duration of time, preventing their interaction with other neurons. This attack does not need previous knowledge by the attacker about the status of the targeted neurons, presenting a low complexity compared to those that could require to study their previous and current status to determine the best instant to attack. 

To formalize this attack, we denote $\mathbb{NE} \subset \mathbb{N}$ as a subset of neurons from the brain, where $n \in \mathbb{NE}$ expresses every single neuron. $\textbf{t}^{\text{attk}}$ is the time instant when the cyberattack starts, and $\textbf{t}^{\text{pulse}}$ is the duration of the attack. During that particular period, a subset of neurons $\mathbb{AN} \subseteq \mathbb{NE}$ is attacked. The voltage of a single neuron in a specific instant of time is denoted as $v_n\in\mathbb{R}$, whereas $v_{min}\in\mathbb{R}$ indicates the minimum value of the voltage that the neuron can have, directly dependent on the neuronal model used in case of simulations. Moreover, $\textbf{t}^{\text{win}}$ is the temporal window in which the cyberattack is evaluated, which corresponds to the duration of the simulation presented in subsequent sections. $\Delta t$ is the amount of time between evaluations during the process, representing the duration of steps of the simulation in the implementation of the cyberattacks. 

As shown in Algorithm \ref{alg:formalJAM}, JAM cyberattacks are performed during a continuous duration of time, where the attacked neurons are forced to have their minimum voltage value. In other words, it avoids the targeted neurons to produce spikes, understood as the inhibition of the neurons.

\begin{algorithm} 
\caption{JAM cyberattack execution} 
\label{alg:formalJAM} 
\begin{algorithmic}
    \STATE $t=0$
    \WHILE{$t < t^{win}$}
        \IF{$t >= t^{attk}$ AND $t < (t^{attk}+t^{pulse})$}
            \FORALL{$n \in \mathbb{AN}$}
                \STATE $v_n \leftarrow v_{min}$
            \ENDFOR
        \ENDIF
        \STATE $t \leftarrow t + \Delta t$
    \ENDWHILE
\end{algorithmic}
\end{algorithm}

\begin{figure}[h]
\begin{center}
\includegraphics[width=\columnwidth]{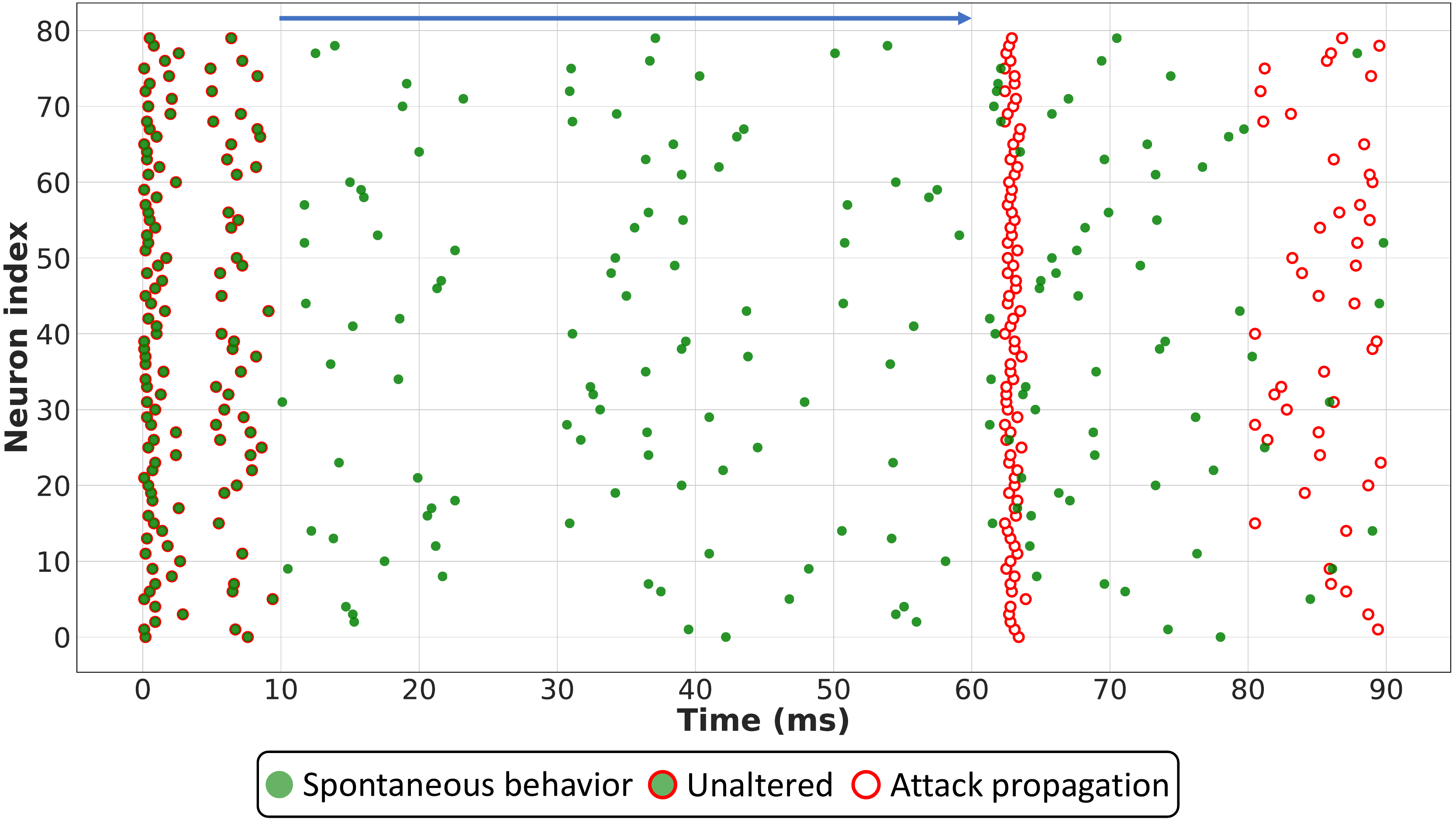}
\end{center}
\caption{Raster plot of a JAM cyberattack when the attack is performed between the instants 10ms and 60ms. This temporal window is represented by a blue arrow for clarity.}
\label{fig:raster_JAM}
\end{figure}

To visually understand the behavior of a JAM cyberattack, \figurename~\ref{fig:raster_JAM} presents the comparison between a JAM cyberattack and the spontaneous neuronal behavior for a simulation of 90ms. Until the instant 10ms, green dots with a red outline can be appreciated, indicating that the attack has not altered those spikes. This attack, performed between the instants 10ms and 60ms, and indicated by a blue arrow, affects all 60 neurons represented in the figure. Because of that, during that temporal window, only green dots are presented, having an absence of neural activity during the application of the attack. After the instant 60ms, white dots with red online appear, which indicate the new spikes generated as a consequence of the attack. It is relevant to note that, from that moment until the instant 90ms, the neural signaling generated by the attack is completely different from the spontaneous behavior.

\section{Experimental Setup}
\label{sec:setup}
Due to the lack of realistic and precise neuronal topologies in the literature, this section presents the methodology followed to create a neuronal topology used to evaluate the impact of JAM cyberattacks. For the sake of simplicity, we have summarized the explanations of this section, where a broader description is available at \cite{Lopez_Bernal:cyberattacks_implants:2020}.

\begin{figure}[h]
\begin{center}
\includegraphics[width=0.7\columnwidth]{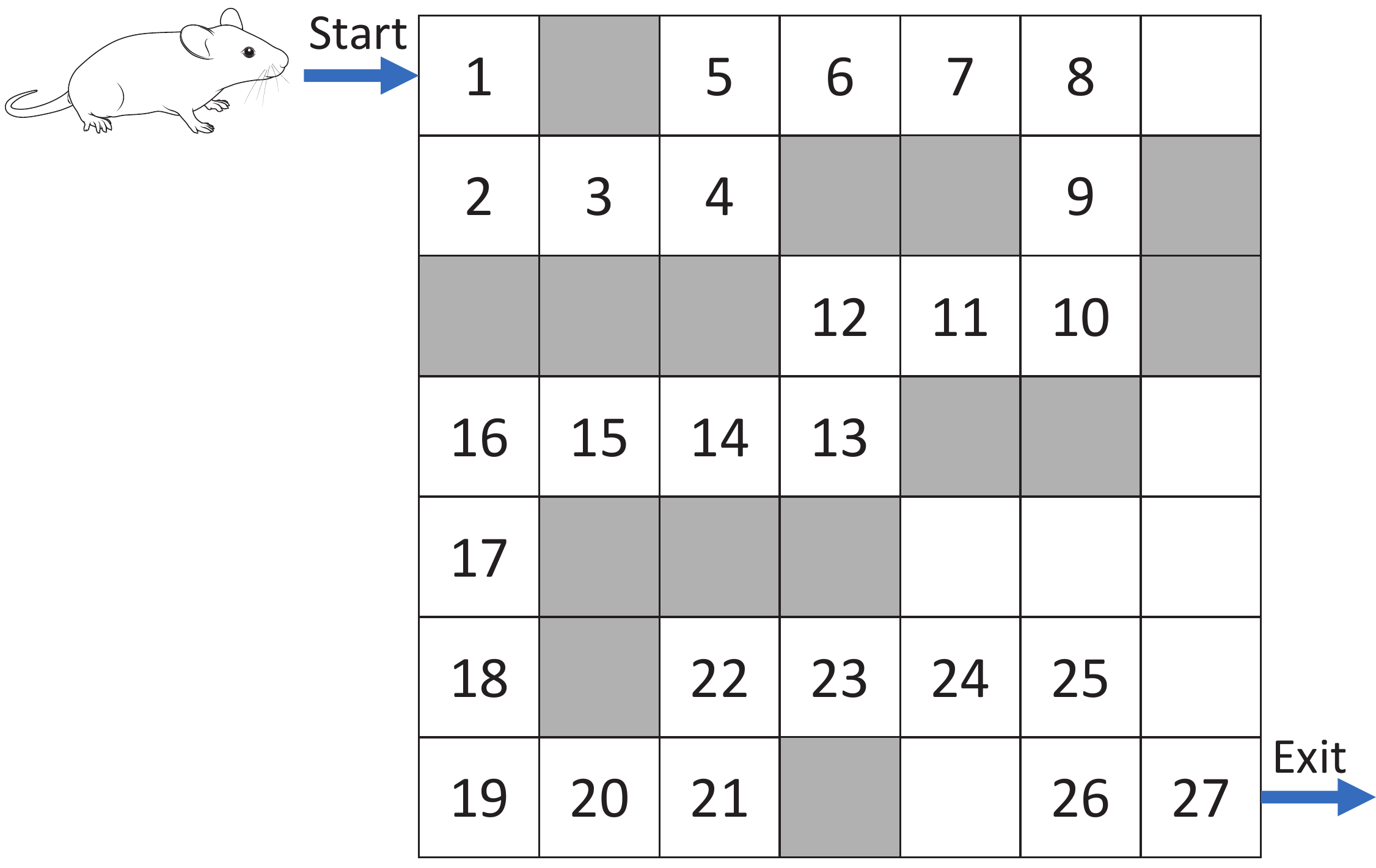}
\end{center}
    \caption{Maze used to model the movement of the mouse, including the optimal path between the starting and final cells.}
\label{fig:mazeLayout}
\end{figure}

It is essential to highlight that the knowledge of precise neocortical synaptic connections in mammalian is nowadays an open challenge \cite{Gal2017}. Although artificial and biological networks cannot be comparable in complexity and functioning, there are works in the literature demonstrating that neurons in the visual cortex present certain similarities with a Convolutional Neural Network (CNN). In this sense, the visual recognition process operates incrementally in both networks, moving from simple to abstract \cite{Kuzovkin2018}. Based of that, we have trained a CNN using Keras on top of TensorFlow \cite{Keras2015} to solve a simplistic scenario based on a mouse trying to escape a maze from any position, inspired in the code from \cite{Zafrany}. The maze has a size of 7x7 positions with fixed obstacles that serve as walls, containing a single starting cell and an exit. \figurename~\ref{fig:mazeLayout} presents the maze, indicating with numbers the optimal path to exit, which has been determined during the training process of the CNN.

The CNN has been trained employing reinforcement learning \cite{Sutton2018}, using a topology consisting in three layers where the first two were convolutional layers, and the third one was dense. After the training process, a topology of interconnected nodes between layers was obtained, where each link had associated a filter weight. These weights represent the relevance that this connection has in the topology to solve the problem.  \tablename~\ref{table:CNN} summarizes the configuration used to define the CNN, composed of a total number of 276 nodes. 

\begin{table}[h]
\caption{Summary of the layers of the CNN}
\label{table:CNN}
\setlength\tabcolsep{2pt}
\resizebox{\columnwidth}{!}{
\begin{tabular}{|c|c|c|c|c|c|c|c|c|}
\hline

Layer & Type & Filters & \begin{tabular}{@{}c@{}}Input \\ size\end{tabular} & \begin{tabular}{@{}c@{}}Output \\ size\end{tabular} & \begin{tabular}{@{}c@{}}Kernel \\ size\end{tabular} & Stride & \begin{tabular}{@{}c@{}}Activation \\ function\end{tabular} & Nodes \\ \hline

1 & Conv2D & 8  & 7$\times$7$\times$1 & 5$\times$5$\times$8 & 3$\times$3 & 1 & ReLU & 200 \\ \hline
2 & Conv2D & 8  & 5$\times$5$\times$8 & 3$\times$3$\times$8 & 3$\times$3 & 1 & ReLU & 72 \\ \hline
3 & Dense  & -  & 3$\times$3$\times$8 & 4 & - & - & ReLU & 4 \\ \hline

\end{tabular}}
\end{table}

The resulting topology was translated to a biological neuronal network by keeping the exact number of layers and nodes per layer and translating the filter weights to synaptic weights. This topology represents a small section of the visual cortex of a mouse. Once having the biological topology, we have used the Brian2 neural simulator \cite{Stimberg2019} to represent the behavior of each individual neuron. In particular, we have implemented the Izhikevich neuronal model \cite{Izhikevich2003}, whose parameters are presented in \tablename~\ref{table:Izhikevich}, and equations (\ref{eq:izhikevich1}), (\ref{eq:izhikevich2}), and (\ref{eq:izhikevich3}). It is relevant to highlight the functioning of the $I$ parameter, used in the experiments to model in the biological simulation the visual stimuli received by the mouse. To enclose the problem, we implemented and monitored a neuronal simulation with a total duration of 27 seconds, where the mouse stayed in one position of the optimal path for one second, and studied its spontaneous behavior and the behavior under attack. When the mouse is in a particular position, the \textit{intervening neurons} associated with each \textit{adjacent position} from the current cell was obtained. The concept of \textit{intervening neurons} can be understood as the set of neurons influenced by the list of adjacent positions from the current cell. For those intervening neurons, the simulation assigns a value of 15mV/ms, keeping a value of 10mV/ms for the rest of the neurons. These particular implementation aspects are presented in-depth in \cite{Lopez_Bernal:cyberattacks_implants:2020}.

\begin{table}[h]
\caption{Parameters used in the Izhikevich model}
\label{table:Izhikevich}
\setlength\tabcolsep{2pt}
\resizebox{\columnwidth}{!}{
\begin{tabular}{|c|c|c|}
\hline

Parameter & Description & Values  \\ \hline

$v$ & Membrane potential of a neuron & [-65, 30] mV \\ \hline
$u$ & Membrane recovery variable providing negative feedback to $v$ & (-16, 2) mV/ms \\ \hline
$a$ & Time scale of $u$ & 0.02/ms \\ \hline
$b$ & Sensitivity of $u$ to the sub-threshold fluctuations of $v$ & 0.2/ms \\ \hline
$c$ & After-spike reset value of $v$ & -65mV \\ \hline
$d$ & After-spike reset value of $u$ & 8mV/ms \\ \hline
$I$ & Injected synaptic currents & \{10, 15\} mV/ms \\ \hline

\end{tabular}}
\end{table}

\noindent
\begin{equation} \label{eq:izhikevich1}
        v'=0.04v^2+5v+140+u+I
\end{equation}

\noindent
\begin{equation} \label{eq:izhikevich2}
        u'=a(bv-u) 
\end{equation}

\noindent
\begin{equation} \label{eq:izhikevich3}
        if v \geqslant 30mV, then \begin{cases}
          v \leftarrow c \\
          u \leftarrow u+d 
        \end{cases}      
\end{equation}

\section{Impact of JAM attacks over biological and artificial neural networks}
\label{sec:resultsJAM}

Once explained the generation of the artificial and biological networks, this section measures and compares the impact generated by Neuronal Jamming cyberattacks (JAM) over biological and artificial networks. In particular, this analysis aims to study if an alteration in neuronal behavior can also impact the mouse's ability to solve the maze.

\tablename~\ref{table:param_JAM_general} presents the parameters used to perform the experiments, indicating between parentheses, if a parameter is common to both scenarios or specific to one of them. As can be seen, five number of simultaneously attacked neurons (named as nodes in the CNN) have been tested, probing several consecutively attacked positions ranging from one to all the positions of the optimal path of the maze. Additionally, each combination of parameters is executed ten times, where each execution selects a different set of randomly selected neurons.

\begin{table}[h]
\caption{Parameters used in the analysis for JAM cyberattacks.}
\label{table:param_JAM_general}
\setlength\tabcolsep{2pt}
\setlength\tabcolsep{1.6pt}
\centering
\resizebox{\columnwidth}{!}{
\begin{tabular}{|c|c|}
\hline

\textbf{Parameter} & \textbf{Values}  \\ \hline

Number of consecutive attacked positions (Bio, CNN) & $\{1, 2, ..., 27\}$ \\ \hline
Number of neurons/nodes (Bio, CNN) & $\{5, 35, 55, 75, 105\}$ \\ \hline
Voltage under attack (Bio) & -65 mV \\ \hline
Output importance (CNN) & -1 \\ \hline
Number of executions (Bio, CNN) & 10 \\ \hline

\end{tabular}}
\end{table}

\subsection{JAM cyberattacks over the biological network}
\label{subsec:JAM_biological}

\begin{figure*}[h]
\begin{center}
\includegraphics[width=0.9\textwidth]{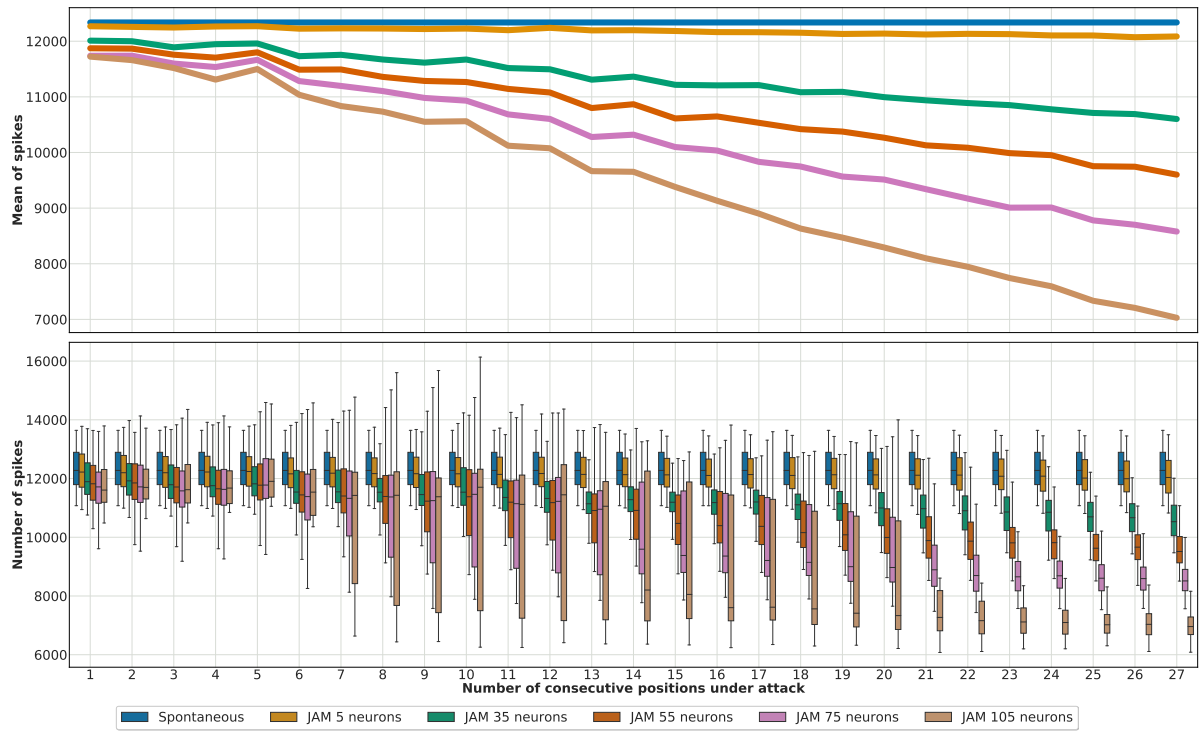}
\end{center}
\caption{Distribution of the number of spikes based on the consecutive number of positions attacked.}
\label{fig:general_number_spikes_JAM}
\end{figure*}

Focusing on the biological perspective, attacked neurons are forced to the minimum voltage value of the model, which corresponds to -65 mV, as indicated in \tablename~\ref{table:param_JAM_general}. \figurename~\ref{fig:general_number_spikes_JAM} presents the experiment consisting of augmenting the number of consecutive positions of the optimal path under attack, always initiating the attack in the first position, and evaluating different numbers of simultaneously attacked neurons. The variability shown corresponds to the ten executions performed per combination of parameters. In particular, this figure highlights how augmenting the number of consecutive positions of the labyrinth under attack impacts in terms of the number of spikes metric. The upper sub-figure depicts that increasing the number of simultaneously attacked neurons generates a considerable reduction in the mean of spikes, reaching a difference of 5000 spikes in the most damaging situation compared to spontaneous behavior. The bottom sub-figure shows that the distribution of the number of spikes presents small variability during the first six positions. More consecutive positions generate a progressive reduction in the dispersion, particularly for higher numbers of attacked neurons, indicating that JAM cyberattacks cause an enormous impact on the spike metric. Nevertheless, increasing the number of consecutive positions over more than 20 generates a progressive reduction in the distributions when attacking more than 75 neurons. This situation is explained by many neurons without activity during the majority of the simulation, decreasing their variability in the number of spikes.

\begin{figure*}[h]
\begin{center}
\includegraphics[width=0.9\textwidth]{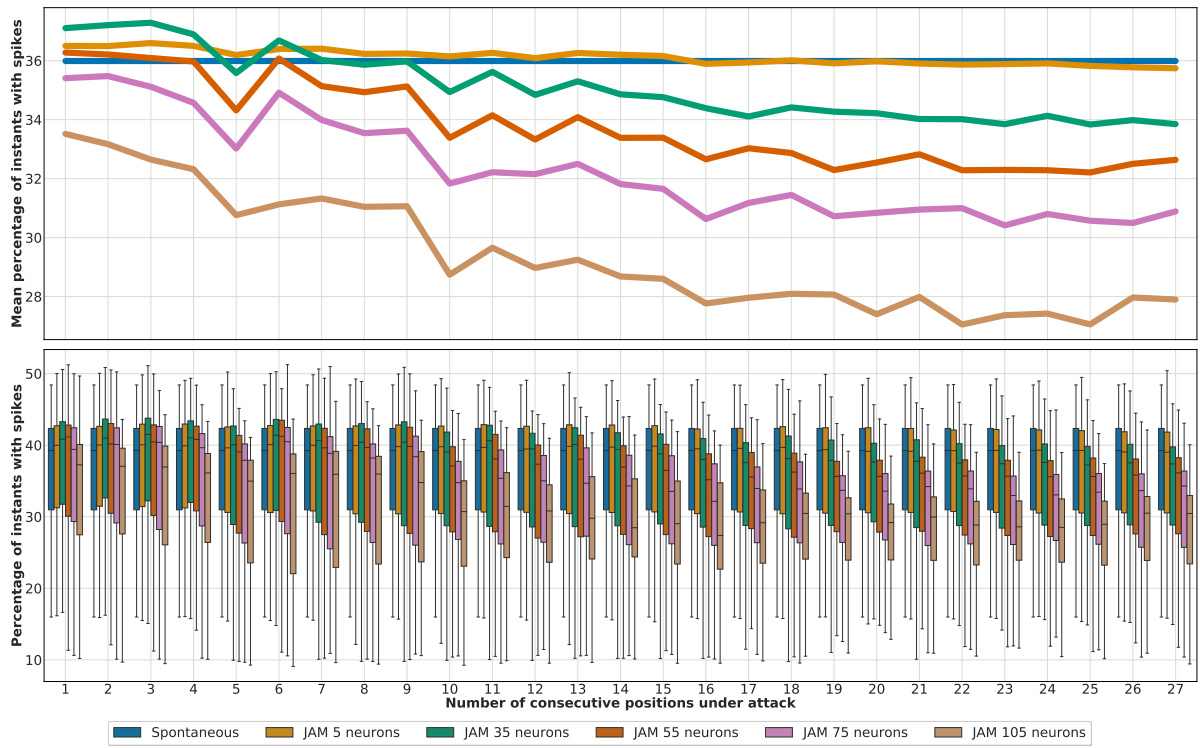}
\end{center}
\caption{Distribution of the temporal dispersion based on the consecutive number of positions attacked.}
\label{fig:general_horizontal_dispersion_JAM}
\end{figure*}

Moving to the temporal dispersion of spikes, \figurename~\ref{fig:general_horizontal_dispersion_JAM} depicts that attacking a higher number of neurons reduces the temporal dispersion. It is relevant to highlight that targeting a reduced number of neurons (up to 35) produces a higher dispersion than the spontaneous behavior. These peaks can be produced by the slight variations generated by the attack. Nevertheless, increasing the number of selected neurons gets a substantial reduction. It is also important to note that, in the bottom sub-figure, the distribution of targeting 105 neurons gets a significant decrease compared to other numbers of attacked neurons, indicating the importance of this parameter of the attack.

\subsection{JAM cyberattacks over the artificial network}
\label{subsec:JAM_artificial}

In the artificial scenario, the attack consists of modifying the targeted nodes of the trained model, affecting their normal functioning. For that, the concept of \textit{output importance} refers to the value used to alter the output of the nodes targeted by the attack, thus affecting their relevance in the network. In JAM, the value used to attack the nodes is -1, which indicates that those nodes do not have any relevance in the network, representing their inhibition. 

\begin{figure}[h]
\begin{center}
\includegraphics[width=\columnwidth]{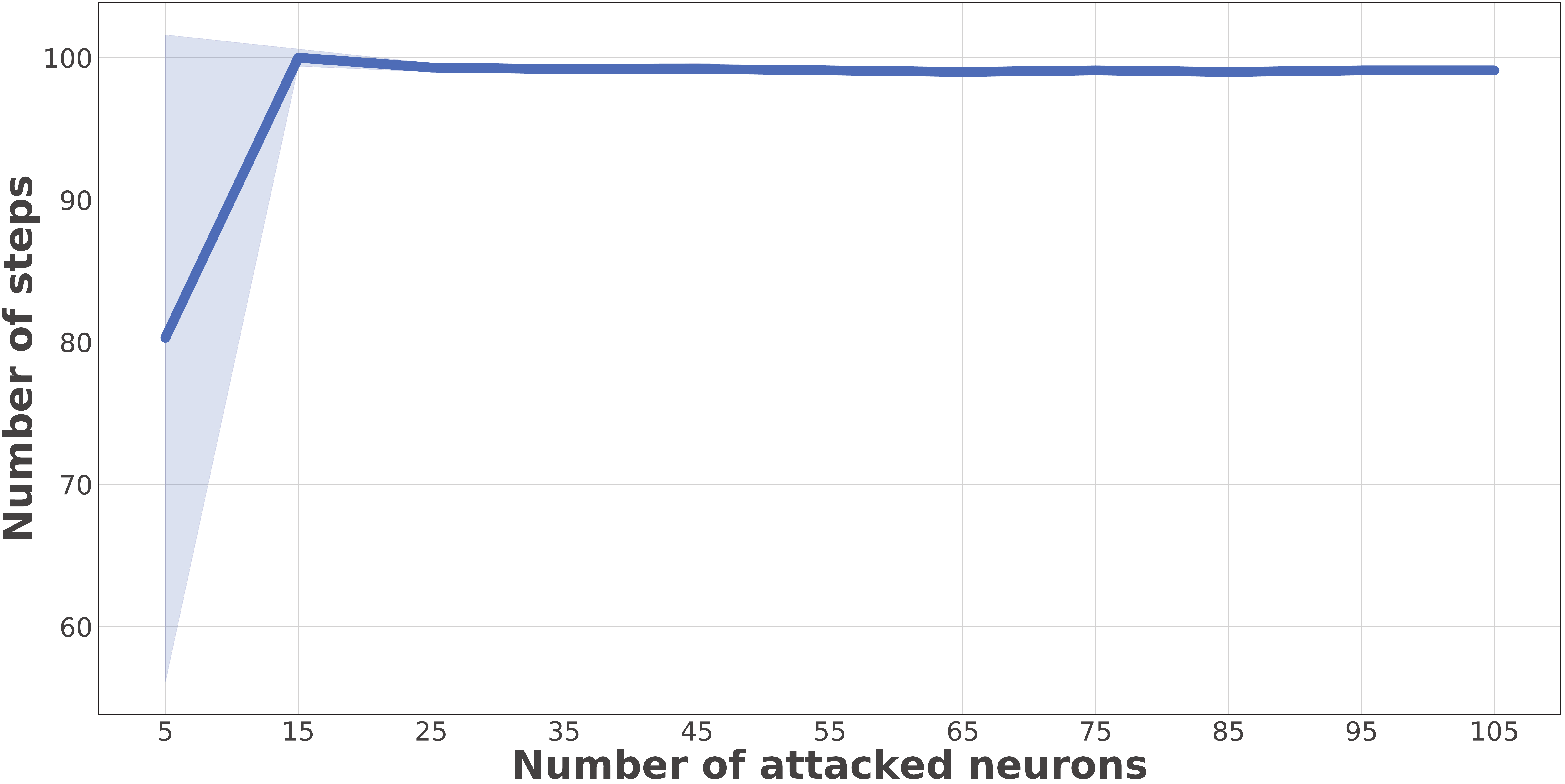}
\end{center}
\caption{Number of steps for different number of neurons between 5 and 105, with 10 executions.}
\label{fig:CNN_always_attack_steps_5-105n}
\end{figure}

The first approach followed was to apply the attacked model for the targeted consecutive positions, restoring it to the non-altered model after the duration of the attack. Although the mouse performed erratic decisions across the maze during the attack, once the model without attacks was restored, the mouse was always able to find the exit position ultimately. To better measure the impact of this attack in terms of percentage of success and number of steps, we decided to continuously perform the attack for all 27 positions of the maze. These experimentation results are represented in \figurename~\ref{fig:CNN_always_attack_steps_5-105n}, which indicates that simultaneously attacking more than 15 nodes does not generate any difference since the number of steps gets constant. It is worthy to note that the success percentage is not studied as both variables are highly correlated. 

\begin{figure}[h]
\begin{center}
\includegraphics[width=\columnwidth]{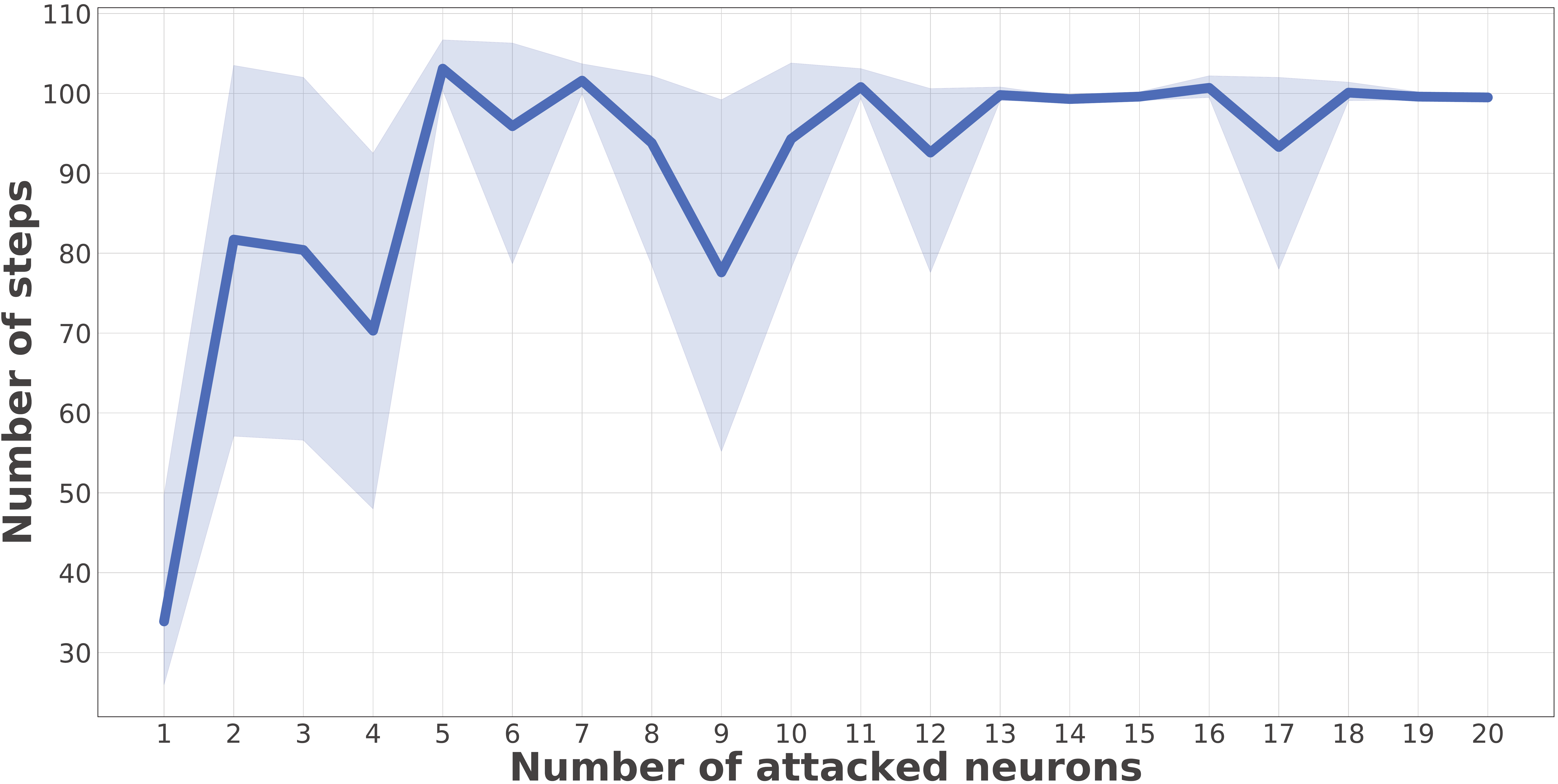}
\end{center}
\caption{Number of steps for a range between one and 20 attacked neurons, with 10 executions.}
\label{fig:CNN_always_attack_steps_1-20n}
\end{figure}

Based on the decision to attack along with all positions, and compare these results with the biological simulation, we decided to focus the analysis of both scenarios on a number of attacked neurons between one and 20. From the CNN point of view, this decision is motivated by \figurename~\ref{fig:CNN_always_attack_steps_1-20n}, which indicates that this particular range reflects variations in the number of steps and that further increments in this variable do not offer new variability. 

\begin{figure}[h]
\begin{center}
\includegraphics[width=\columnwidth]{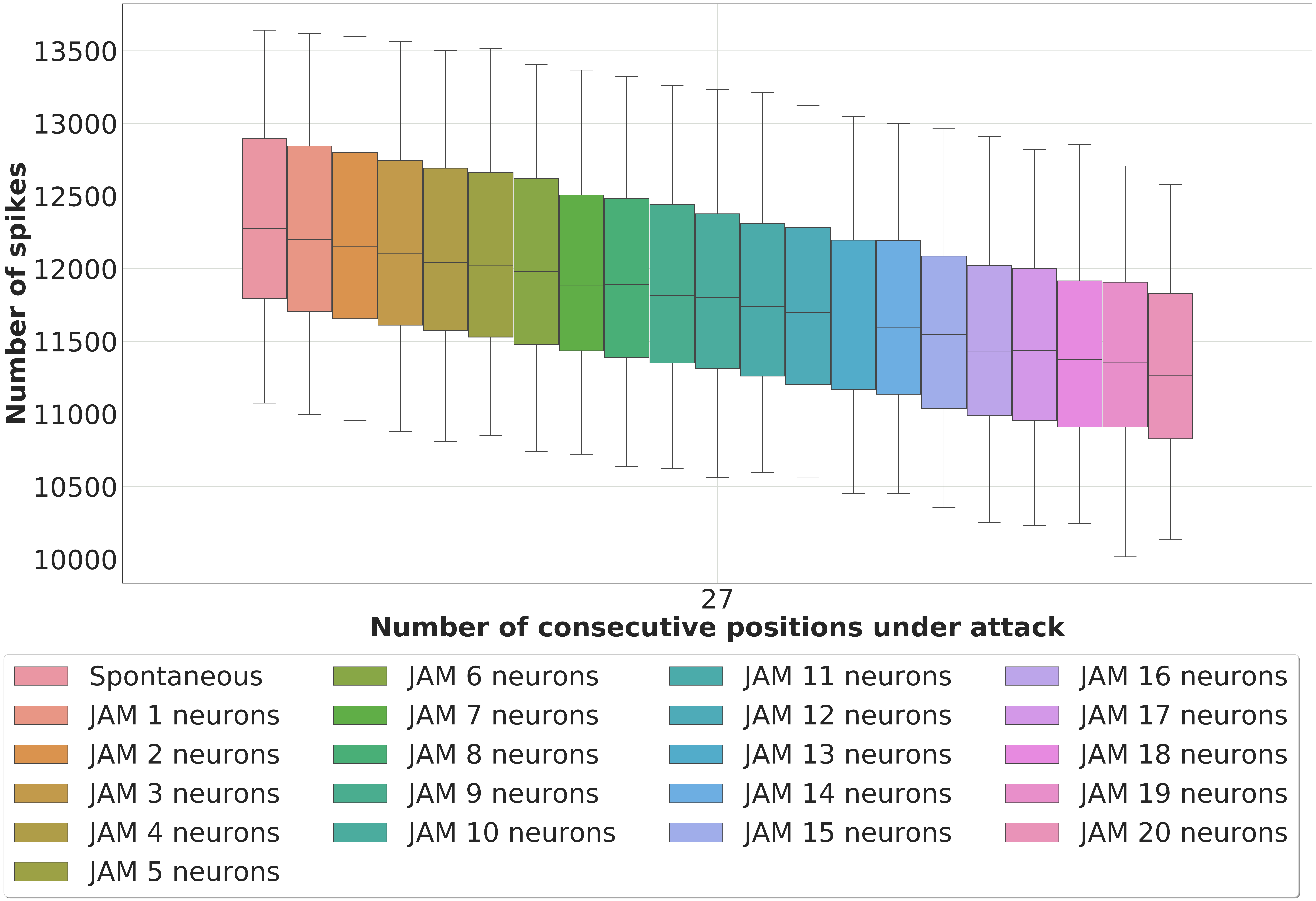}
\end{center}
\caption{Number of spikes for a range of attacked neurons between 1 and 20.}
\label{fig:number_spikes_1-20n}
\end{figure}

\begin{figure}[h]
\begin{center}
\includegraphics[width=\columnwidth]{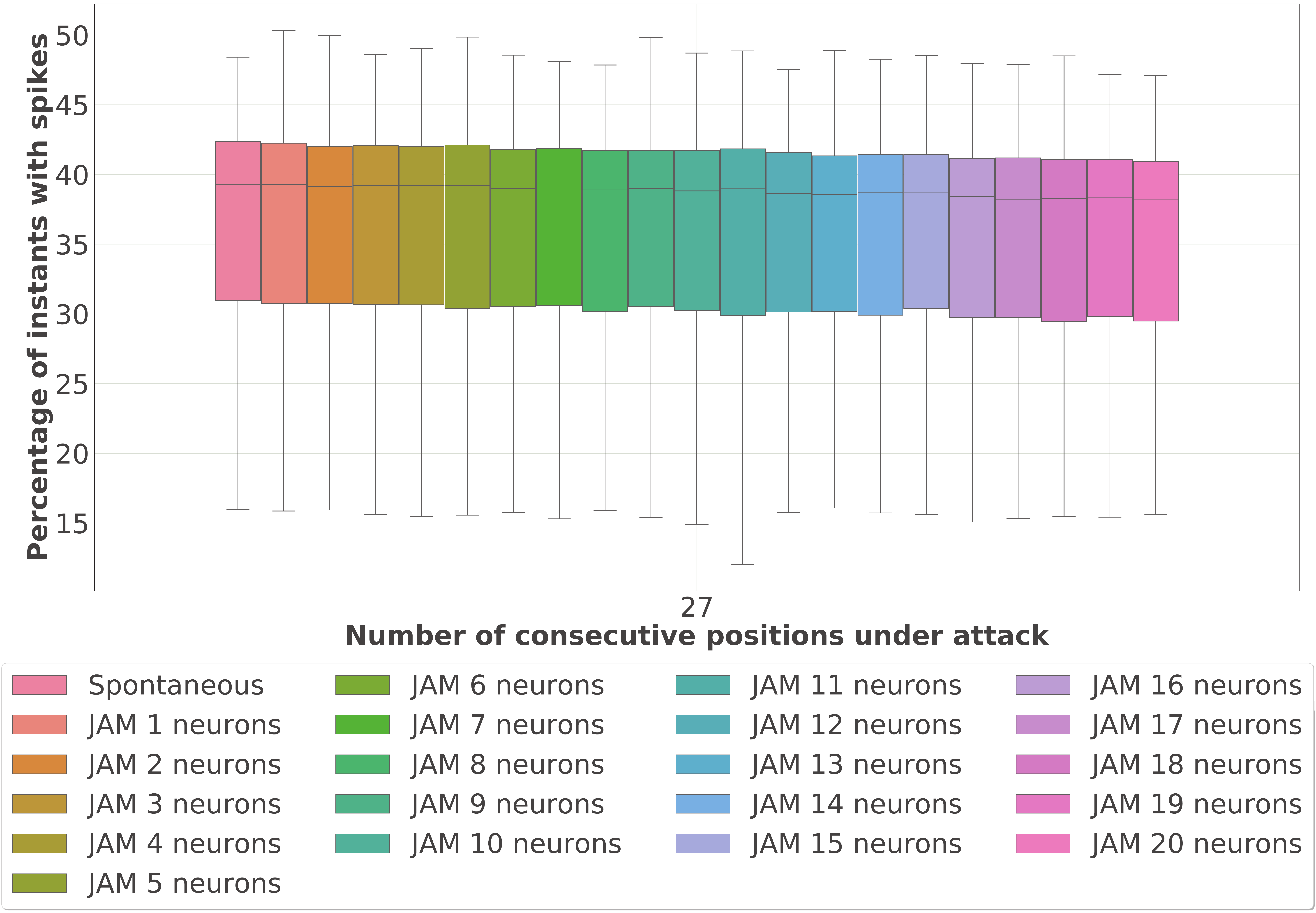}
\end{center}
\caption{Temporal dispersion for a range of attacked neurons between 1 and 20.}
\label{fig:horizontal_dispersion_1-20n}
\end{figure}

After defining the range, the biological experiments were adapted to be comparable with those from the CNN scenario. For that, a number of attacked neurons between one and 20 were selected, setting the attack to cover all 27 consecutive positions of the optimal path of the maze, starting in the instant 50ms. \figurename~\ref{fig:number_spikes_1-20n} and \figurename~\ref{fig:horizontal_dispersion_1-20n} present, respectively, the results for the number of spikes and the temporal dispersion. It is important to highlight that these plots present the same trend as described in \figurename~\ref{fig:general_number_spikes_JAM} and \figurename~\ref{fig:general_horizontal_dispersion_JAM}, respectively, for the analysis between 5 and 105 attacked neurons. 

Finally, \tablename~\ref{table:corr_JAM} compares the correlation between both scenarios, which determines a correlation between the number of steps and the number of spikes of -0.66, indicating a specific relationship between both metrics. A similar situation happens between the number of steps and the percentage of dispersion, with a -0.59 Pearson correlation. This indicates, in general, a low correlation between scenarios. However, this can be explained due to the reduction in the number of attacked neurons considered. As indicated before, the number of neurons has been limited to a range between one and 10. Although these values offer variability in the CNN, there is not much difference in the distribution between these close sizes in the biological simulation. Nevertheless, the individual analysis performed in this section for both biological and artificial scenarios presents the high impacts that JAM cyberattacks generate over these scenarios.

\begin{table}[h]
\caption{Correlation of relevant features between CNN and biological experiments for JAM attacks.}
\label{table:corr_JAM}
\setlength\tabcolsep{2pt}
\centering
\begin{tabular}{|c|c|c|c|c|}
\hline

&  \# spikes & \% dispersion & \# steps & \# neurons \\ \hline
\# spikes & 1.00 & 0.98 & -0.66 & -0.99 \\ \hline
\% dispersion & 0.98 & 1.00 & -0.59 & -0.98 \\ \hline
\# steps & -0.66 & -0.59 &  1.00 & 0.66  \\ \hline
\# neurons & -0.99 & -0.98 & 0.66 & 1.00   \\ \hline

\end{tabular}
\end{table}

\section{Comparison of JAM and FLO cyberattacks}
\label{sec:resultsFLO}

This section compares the impact caused by JAM cyberattacks with FLO, a neuronal cyberattack existing in the literature. For that, we first introduce FLO cyberattacks, moving to the analysis of their impacts, and later we compare it with JAM. This section also provides an in-depth study of the results of individually performing FLO cyberattacks in different positions of the optimal path, comparing the results of biological and artificial networks.

\subsection{Definition of Neuronal Flooding cyberattacks}

Neuronal Flooding cyberattacks (FLO) were defined in our previous work \cite{Lopez_Bernal:cyberattacks_implants:2020} as a way to overstimulate targeted neurons, In that work, we just explored the cyberattacks for the first position of the maze, behavior that is formally represented by Algorithm \ref{alg:formalFLO}. In particular, it indicates that the attack over the targeted neurons is performed in a particular instant of time $\textbf{t}^{\text{attk}}$, in contrast to JAM, which is executed within a determined temporal period.

\begin{algorithm} 
\caption{FLO cyberattack execution} 
\label{alg:formalFLO} 
\begin{algorithmic}
    \STATE $t=0$
    \WHILE{$t < t^{win}$}
        \IF{$t == t^{attk}$}
            \FORALL{$n \in \mathbb{AN}$}
                \STATE $v_n \leftarrow v_n+vi_n$
            \ENDFOR
        \ENDIF
        \STATE $t \leftarrow t + \Delta t$
    \ENDWHILE
\end{algorithmic}
\end{algorithm}

In contrast, the current work performs attacks over each individual position of the optimal maze path, evaluating a different number of simultaneously attacked neurons and multiple increment voltages per position. The parameters used for this experiment are indicated in \tablename~\ref{table:param_FLO_general}, having five different values of simultaneously attacked neurons (or nodes) and four different voltage increments. Besides, each combination of parameters is executed ten times. 

\begin{table}[h]
\caption{Parameters used in the analysis for FLO cyberattacks.}
\label{table:param_FLO_general}
\setlength\tabcolsep{2pt}
\centering
\begin{tabular}{|c|c|}
\hline

\textbf{Parameter} & \textbf{Values}  \\ \hline

Positions attacked (Bio, CNN) & $\{1, 2, ..., 27\}$ \\ \hline
Number of neurons/nodes (Bio, CNN) & $\{5, 35, 55, 75, 105\}$ \\ \hline
Voltage increment (Bio) & [10, 20, 40, 60] mV \\ \hline
Output importance (CNN) & [15, 30, 60, 90] \% \\ \hline
Number of executions (Bio, CNN) & 10 \\ \hline

\end{tabular}
\end{table}

\subsection{FLO cyberattacks over the biological network}
\label{subsec:FLO_biological}

In the biological scenario, we perform a FLO cyberattack 50ms after the mouse reaches each position of the maze, evaluating the impact of the attack during the complete simulation (27 seconds, until the mouse reaches the exit) based on the number of spikes and temporal dispersion metrics.

\begin{figure*}[h]
\begin{center}
\includegraphics[width=0.9\textwidth]{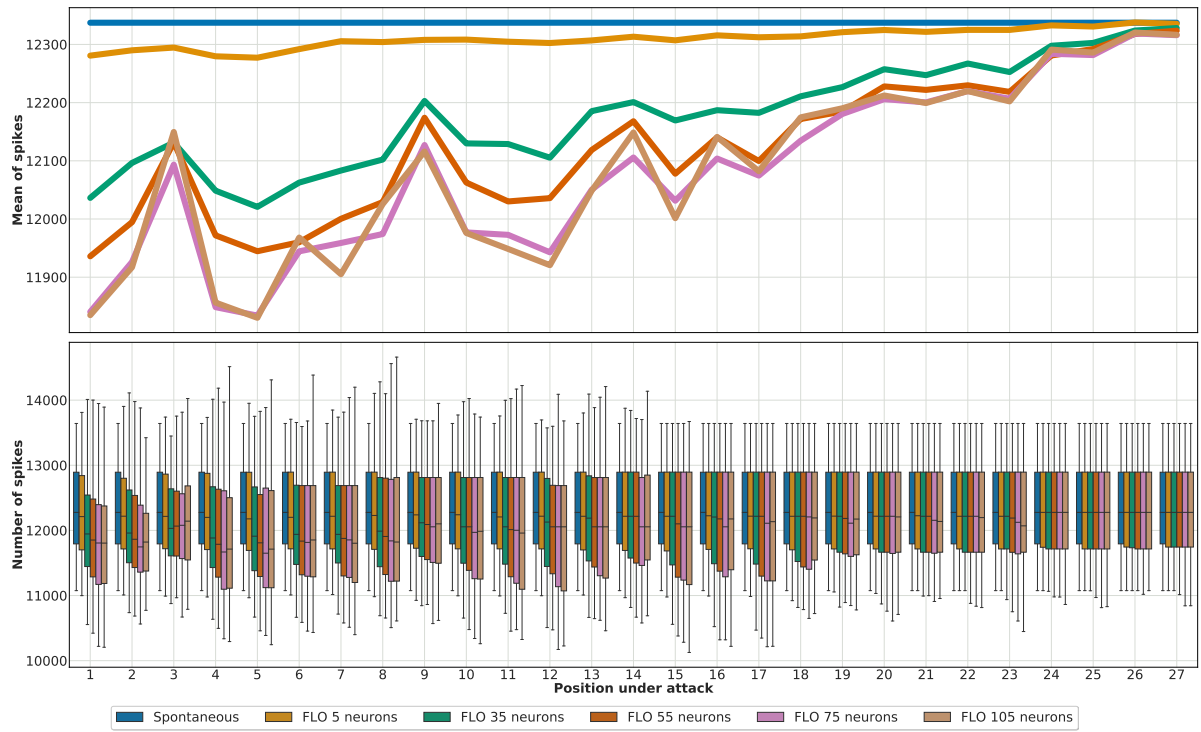}
\end{center}
\caption{Distribution of the number of spikes according to the position under attack.}
\label{fig:general_number_spikes_FLO}
\end{figure*}

\figurename~\ref{fig:general_number_spikes_FLO} presents the evolution of the number of spikes according to the individual position of the optimal path under attack, fixing the voltage increment to a value of 40 mV for the sake of simplicity. It is worthy to note that, for each attacked position, the represented values correspond to the number of spikes over the complete simulation. The upper sub-figure presents the mean of spikes per attacked position, where each line represents a different number of attacked neurons. It also indicates that this attack causes a desynchronization of neuronal activity over time, presenting a higher dispersion when the attack is performed in the first positions. This dispersion is also benefited by the particular model used and the propagation of the spikes. Since FLO cyberattacks over neuronal activity cause a reduction in the number of spikes \cite{Lopez_Bernal:cyberattacks_implants:2020}, performing the attack in lasts positions means that the previous positions of the optimal path were not modified by the attack, being the same as the spontaneous behavior. 

Additionally, attacking a broader number of neurons produces, in general, a higher reduction in the mean of spikes. Nevertheless, there are no significant differences between attacking 75 and 105 simultaneous neurons. These data correspond to the mean of the distribution represented in the bottom sub-figure, where we can see a higher variability in the number of spikes when the attack is applied in the first positions. This figure also highlights that the maximum and minimum values of the distribution have a significant variability compared to the spontaneous behavior, stabilized when we attack in later positions.

\begin{figure*}[h]
\begin{center}
\includegraphics[width=0.9\textwidth]{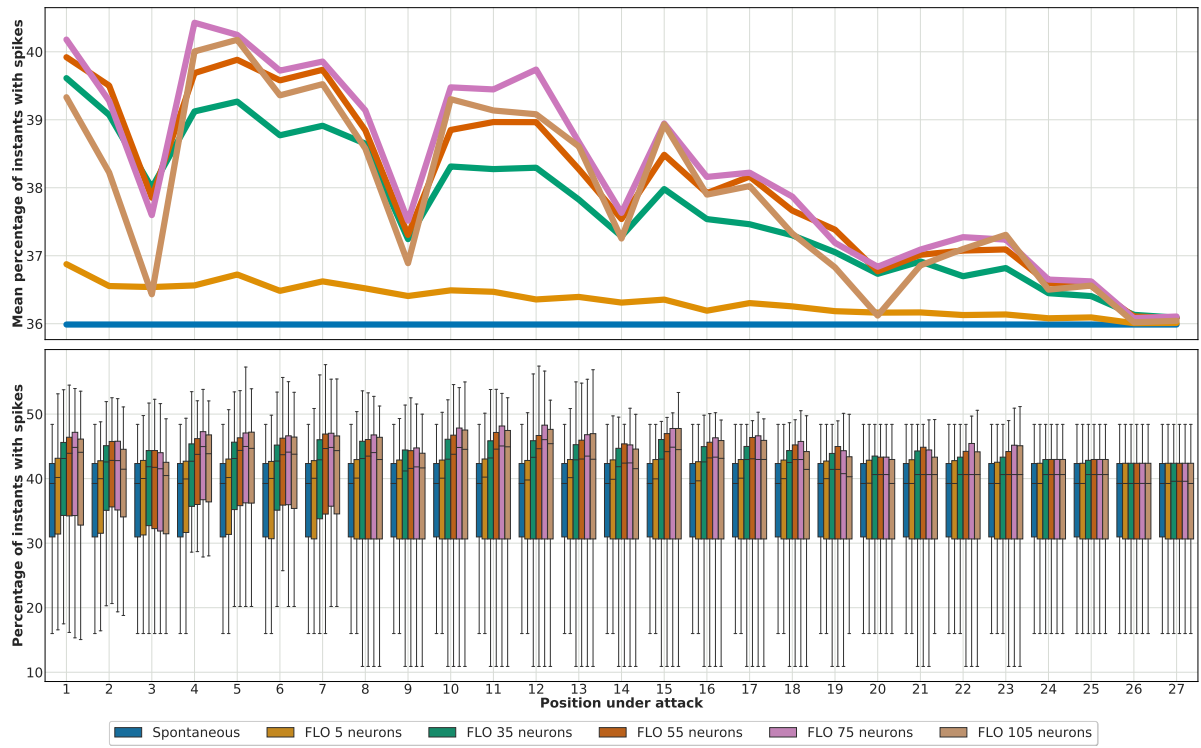}
\end{center}
\caption{Percentage of instants with spikes according to the position under attack.}
\label{fig:general_horizontal_dispersion_FLO}
\end{figure*}

After analyzing the behavior of the FLO cyberattack in terms of the number of spikes, \figurename~\ref{fig:general_horizontal_dispersion_FLO} presents its impact focusing on the temporal dispersion metric. As can be seen, the dispersion is higher when attacking the first positions due to the same reasons addressed for the number of spikes metric. Additionally, attacking a broader number of neurons derives in a higher percentage of instants with spikes. Finally, it is worthy to note that these two metrics are highly related, with Pearson correlation value of -0.97.

\subsection{FLO cyberattacks over the artificial network}
\label{subsec:FLO_artificial}

In terms of attacks over the CNN, it is essential to note that the voltage increments used to attack the biological network have been proportionally adapted to the CNN scenario, corresponding to the output importance indicated in Table~\ref{table:param_FLO_general}. Based on that, and as an example, a value of $10mV$ in the biological scenario represents a 15\% from the voltage range defined by the Izhikevich model used. This 15\% is the equivalent value used to increment the importance of the targeted nodes during the attack to the CNN. The rest of the output importance values are calculated the same way.

\begin{figure}[h]
\begin{center}
\includegraphics[width=\columnwidth]{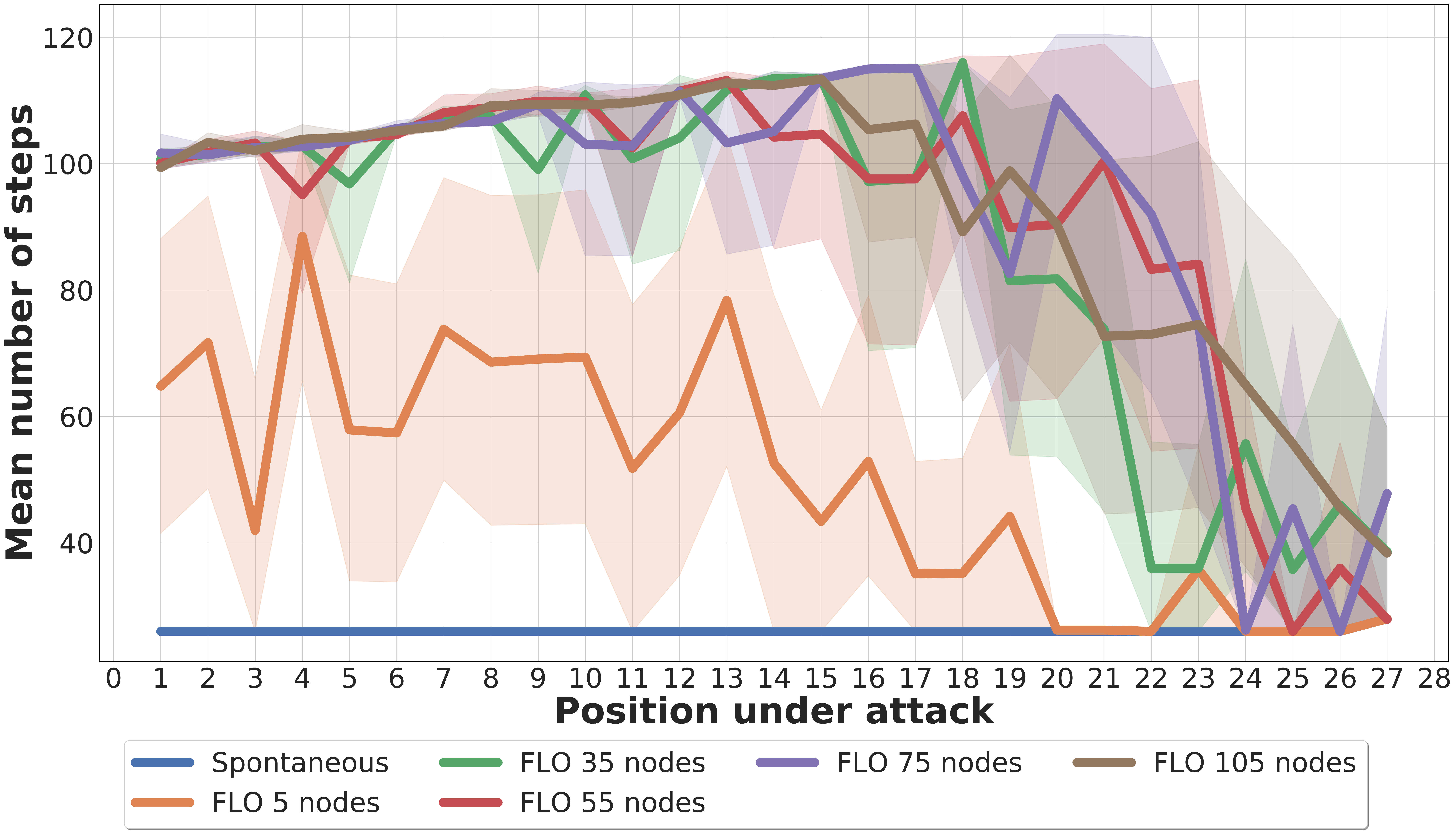}
\end{center}
\caption{Mean of steps when we perform an attack in each position of the optimal path of the maze, considering five different number of simultaneously attacked neurons.}
\label{fig:general_number_steps_FLO}
\end{figure}

\figurename~\ref{fig:general_number_steps_FLO} presents the evolution of the mean number of steps among the ten executions per combination of parameters. For each attacked position, the mouse was located in that position, where the attack was performed to alter the model for the targeted neurons, consequently evaluating the number of steps required to exit the maze from the starting position of the labyrinth. It is essential to note that, once the model is attacked, it is used until the end of that particular execution. In this figure, each color indicates a different number of simultaneous neurons attacked. It can be appreciated that the number of steps remains constant in the spontaneous behavior of the CNN, requiring 26 steps to find the exit. There is an exception in position 27, where the mouse needs to move to an adjacent cell in the maze to finally reach the exit since the mouse initially started in the exit position. This figure highlights that augmenting the number of attacked neurons generates an increase in the number of steps until position 20. From that position, the trend decreases since the closer the mouse is to the exit, the easier it is to solve the maze by probability.

Another relevant metric is the percentage of times in which the mouse finds the exit. The Pearson correlation has been calculated between the number of steps and the success rate, obtaining a value of -0.99, meaning that they present a trend almost identical in an inversely proportional way. That is to say, the number of steps increases when the percentage of success gets reduced. Thus, the number of steps will be the sole metric used to evaluate the CNN in this analysis. 

It is interesting to consider the relationship between the results obtained from attacking the biological and artificial scenarios to help understand the behavior in the biological network. To perform this comparison, \tablename~\ref{table:corr_FLO_general} presents the Pearson correlation between the relevant features considered in these domains. In particular, we are interested in the relationship between the number of steps and the number of spikes, and the percentage of dispersion. Based on that, it can be determined that the CNN and biological approaches have a high correlation, with an approximate 80\% correlation in both of them. 

\begin{table}[h]
\caption{Correlation of relevant features between CNN and biological experiments.}
\label{table:corr_FLO_general}
\setlength\tabcolsep{2pt}
\setlength\tabcolsep{1.6pt}
\centering
\resizebox{\columnwidth}{!}{
\begin{tabular}{|c|c|c|c|c|c|}
\hline

 & position of attack & \# spikes & \% dispersion & \# steps & \# neurons \\ \hline
position attack & 1.00  & 0.53  & -0.53 & -0.42 & -0.0  \\ \hline
\# spikes & 0.53 & 1.00  & -0.97 & -0.82 & -0.66 \\ \hline
\% dispersion & -0.53 & -0.97 & 1.00  & 0.81  & 0.56  \\ \hline
\# steps & -0.42 & -0.82 & 0.81  & 1.00  & 0.65  \\ \hline
\# neurons & -0.0  & -0.66 & 0.56  & 0.65  & 1.00  \\ \hline

\end{tabular}}
\end{table}

Based on the above, we can conclude that there is a significant relationship between the results obtained in both experimental dimensions. These results suggest that performing attacks over the brain of the mouse could not only alter its spontaneous neuronal behavior but also affect its decisions to solve the maze. Nevertheless, these results are limited to our use case, the neuronal topology, and the use of a CNN to model a portion of the mouse's visual cortex.

Once presented the relationship between the biological and artificial scenarios, this section compares the results of both attacks. Since the approaches followed between these attacks are not directly comparable, where FLO focuses on individually attacking different positions, and JAM affects multiple consecutive positions, this study focuses on analyzing the correlations obtained for each attack. In FLO, the Pearson correlation obtained was -0.82 for the relationship between the number of steps and number of spikes, and 0.81 between steps and temporal dispersion. On the contrary, a value of -0.66 was obtained between the steps and the spikes and -0.59 for the relationship between steps and dispersion for JAM. These values indicate that the relationship between the biological and artificial networks is closer in the FLO situation, despite the analysis for the JAM cyberattack presented some limitations as stated in Section~\ref{sec:resultsJAM}.

\section{Discussion}
\label{sec:discussion}

This section discusses the results obtained in this work, aiming to understand better the impact of these attacks, their possible consequences in the real world, and defend against them. Additionally, if we could reproduce the effect of neurodegenerative diseases with these attacks, we could generate databases containing multiple attack configurations, study their impact, and propose mechanisms to reduce these impacts.

Previous sections have highlighted the enormous impact that neuronal cyberattacks can cause over spontaneous neural activity, affecting the amount, periodicity, and even the presence of spikes. Additionally, we have observed that these cyberattacks could also alter the simulated mouse's decision ability, forcing it to make mistakes in the resolution of the labyrinth. These cyberattacks possess differences based on their action mechanisms. JAM cyberattacks focus on continuously inhibiting the neuronal activity of the targeted neurons, suppressing this signaling along with the duration of the attack. On the contrary, FLO cyberattacks aim to overstimulate a set of neurons in a particular instant, extending its impact after its application.

Based on these action mechanisms, we identify that the behavior of the previous attacks has similarities with the effects and consequences that certain neurodegenerative diseases generate. As indicated in Section~\ref{sec:related}, neurodegenerative diseases can be included within the concept of brain connectivity disorders. In particular, for Alzheimer's Disease (AD), the deactivation of the Default Mode Network (DMN) could be reproduced by an attacker able to target individual neurons, reproducing or accelerating the effects of the disease. We identify that JAM, focused on neuronal activity inhibition, could be used for these purposes. On the contrary, Amyotrophic lateral sclerosis (ALS) is based on neuronal activity hyperexcitability, where FLO could be applied to periodically stimulate the targeted neurons and thus produce a perturbation in the excitatory/inhibitory balance of cortical neurons.

Although neuronal cyberattacks are promising mechanism aiming to extend our knowledge about cybersecurity on BCI, further research is required to study the impact these cyberattacks can cause over neural circuits and cognitive and behavioral functions. The study of neuronal cyberattacks could help identify particular characteristics helping to detect prospect threats on BCI systems. Additionally, the application of neuronal cyberattacks could be beneficial in neurological research, using these cyberattacks to control the spread of the disease in neural models or even in vivo trials.

\section{Conclusion}
\label{sec:conclusion}

This work introduces the Neuronal Jamming cyberattack (JAM), consisting in the inhibition of neuronal activity. To implement this attack, and due to a lack of realistic neuronal topologies, a Convolutional Neural Network (CNN) has been trained to generate a neuronal topology based on a use case of a mouse trying to exit a maze. Once having both topologies, we analyze the impact that JAM cyberattacks present over biological and artificial scenarios. Additionally, this manuscript offers a comparison between JAM and FLO cyberattacks. For that, we have implemented several configurations of FLO, a cyberattack already existing in the literature aiming to overstimulate neural activity. To measure their impact, we have studied multiple metrics in the biological scenario (number of spikes and temporal dispersion) and in the CNN (number of steps and success rate in solving the problem).

The obtained results highlight that, in JAM cyberattacks, increasing the number of consecutive positions under attack produces a reduction in spikes and temporal dispersion. In the artificial network, attacking up to 20 neurons is enough to prevent the mouse from completing the labyrinth. The comparison between scenarios indicates that this attack could affect the mouse's ability to escape the maze. We have obtained a Pearson's correlation of 0.6, a low value explained due to the restriction of the number of neurons used to compute the correlations. Additionally, we have observed for FLO experiments that delaying the instant of attack to later positions reduces the impact from both biological metrics, generating an increase in the number of steps. Pearson's correlation between variables for this cyberattack was approximately 0.8, highlighting a closer relationship between scenarios. Finally, we have discussed the similarities between neurodegenerative diseases and the neuronal cyberattacks studied. 

In future work, we plan to investigate new neuronal cyberattacks with different action mechanisms and impacts. Additionally, we aim to explore the possibility of having realistic topologies, which are currently very limited, to simulate existing and prospecting cyberattacks. Finally, we want to focus our efforts on designing and implementing detection mechanisms to identify the initiation of a neuronal cyberattack and propose mitigation techniques to reduce their impact or even suppress it. 

\section*{Declaration of Competing Interest}
The authors declare that they have no known competing financial interests or personal relationships that could have appeared to influence the work reported in this paper. 

\section*{CRediT authorship contribution statement}

\textbf{Sergio L\'opez Bernal.} Methodology, Writing - original draft, Data curation, Software. 
\textbf{Alberto Huertas Celdr\'an.} Methodology, Conceptualization, Writing - Review \& Editing.
\textbf{Gregorio Mart\'inez P\'erez.} Supervision, Project administration, Funding acquisition. 

\section*{Acknowledgment}

This work has been partially supported by Armasuisse S+T with project CYD-C-2020003.

%
%
\bibliographystyle{cas-model2-names}
\bibliography{main}

\newpage

\bio{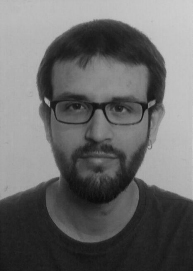}Sergio L\'opez Bernal received the B.Sc. and M.Sc. degrees in computer science from the University of Murcia, and the M.Sc. degree in architecture and engineering for the IoT from IMT Atlantique, France. He is currently pursuing the Ph.D. degree with the University of Murcia. His research interests include ICT security on brain–computer interfaces and network and information security.\endbio

\bio{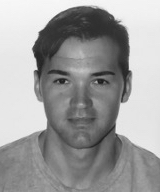}Alberto Huertas Celdr\'an received the M.Sc. and Ph.D. degrees in computer science from the University of Murcia, Spain. He is currently a postdoctoral fellow associated with the Communication Systems Group (CSG) at the University of Zurich UZH. His scientific interests include medical cyber-physical systems (MCPS), brain–computer interfaces (BCI), cybersecurity, data privacy, continuous authentication, semantic technology, context-aware systems, and computer networks.\endbio

\bio{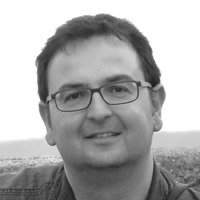}Gregorio Mart\'inez P\'erez is Full Professor in the Department of Information and Communications Engineering of the University of Murcia, Spain. His scientific activity is mainly devoted to cybersecurity and networking, also working on the design and autonomic monitoring of real-time and critical applications and systems. He is working on different national (14 in the last decade) and European IST research projects (11 in the last decade) related to these topics, being Principal Investigator in most of them. He has published 160+ papers in national and international conference proceedings, magazines and journals.\endbio

\end{document}